\documentclass[dvips,12pt]{article}

\usepackage{palatino,amsmath}

\newcommand {\half} {{1 \over 2}}

\newcommand {\ket}[1] {\left| #1 \right>}
\newcommand {\braket}[2] {\left< { \left. #1 \,\right|} \,#2 \right>}

\newcommand {\im} {\mathrm{im}\,}
\newcommand {\sig} {\mathrm{sig}\,}
\newcommand {\Lmparrow} {\genfrac{}{}{0pt}{1}{\xrightarrow{\phantom{n}L_-\phantom{n}}}
                         {\xleftarrow[\phantom{n}L_+\phantom{n}]{}}}

\begin{document}

    \title{Cohomology and Decomposition of
    Tensor Product Representations of $SL(2,\mathbf{R})$}
    \author{Andr\'e van Tonder
            \\ \\
            Department of Physics, Brown University \\
            Box 1843, Providence, RI 02906 \\
            andre@het.brown.edu
            }
    \date{December 12, 2002  \\Revised October 9, 2003}

    \maketitle

    \begin{abstract}
        \noindent
        We analyze the decomposition of tensor products
        between infinite dimensional (unitary)
        and finite-dimensional (non-unitary)
        representations of $SL(2,\mathbf{R})$.
        Using classical results on indefinite inner product spaces,
        we derive explicit decomposition formulae, true modulo a
        natural
        cohomological reduction, for the tensor products.

        \noindent
        \rule{4.6in}{1pt}
        \newline
        Brown preprint: BROWN-HET-1338
    \end{abstract}

    \section{Introduction}

    The representation theory of $SL(2,\mathbf{R})$ was developed in the
    references listed in \cite{SUONEONE}.
    Since $SL(2,\mathbf{R})$ is non-compact, its unitary irreducible representations
    are infinite-dimensional and fall into several classes: the
    principal discrete series, the principal continuous series and
    the exceptional series.  There is also a series of
    finite-dimensional representations which are non-unitary.
    For a recent example of the application of both these unitary and
    non-unitary representations in physics, see
    \cite{BALASU}.

    While an exhaustive analysis of the decomposition of tensor products
    between
    unitary (infinite dimensional) representations
    was carried out by the authors listed in
    reference \cite{DECOMP}, they did not address the coupling of unitary and
    non-unitary representations.

    In this paper we analyze the decomposition of tensor products
    between infinite dimensional (unitary)
    and finite dimensional (non-unitary)
    representations of $SL(2,\mathbf{R})$.
    We show that in the cases where these tensor products are not
    completely reducible,
    there exists a cohomological reduction of the product
        representation
        with respect to a nilpotent operator
        constructed from the Casimir.  On
    this cohomology, the product representations
    become completely reducible.
    Using classical results on indefinite inner product spaces,
        we derive explicit decomposition formulae
        for the tensor products.

    While several authors have discussed aspects of the
    decomposition of tensor products of finite-dimensional and
    infinite-dimensional representations of Lie groups
    \cite{FINTIMESINF},
    to our knowledge the methods and results of the current paper
    are new.

    \section{Representations of $SL(2,\mathbf{R})$}
    \label{REPRESENTATIONS}

    In this section we summarize some basic facts regarding
    the representation theory of $SL(2,\mathbf{R})$ (see \cite{SUONEONE} and
    \cite{BALASU}).

    The commutation relations of the $SL(2,\mathbf{R})$ Lie algebra are
    \[
        [L_1, L_2] = -iL_0,\quad [L_2, L_0] = iL_1,\quad [L_0, L_1] =
        iL_2.
    \]
    Given a representation of $SL(2,\mathbf{R})$, we can define ladder operators
    \begin{eqnarray}
        L_- &\equiv &L_1 + iL_2, \nonumber \\
        L_+ &\equiv &L_1 - iL_2  \label{LADDER}
    \end{eqnarray}
    satisfying
    \[
        [L_+, L_-] = 2 L_0,\quad [L_0, L_\pm] = \mp L_\pm.
    \]
    The operator $L_-$ ($L_+$) increases (decreases) the $L_0$
    eigenvalue $m$ (called the weight) by one unit.
    A state annihilated by $L_+$ is conventionally called highest weight and
    a state annihilated by $L_-$ lowest weight.
    Representations of $SL(2,\mathbf{R})$ are constructed by starting from
    some eigenstate of $L_0$ and repeatedly acting on it with $L_\pm$.
    The chain of states obtained in this way does not necessarily
    terminate on a highest or lowest weight state.

    The Casimir operator is
    \begin{eqnarray}
        \mathbf{L}^2 &=& - L_1^2 - L_2^2 + L_0^2 \nonumber \\
        &=& - L_- L_+ + {L_0} (L_0 -1) \nonumber \\
        &=& - L_+ L_- + {L_0} (L_0 + 1).
        \label{CASIMIR}
    \end{eqnarray}
    The unitary irreducible representations of $SL(2,\mathbf{R})$ are all
    infinite-dimensional and are labelled
    by the parameter $h$ defined through
    $$\mathbf{L}^2 = h(h-1),$$
    as well as by the parameter $\epsilon \in \{0, \half\}$
    which determines whether the spectrum of $L_0$ is integral or
    half-integral.

    The following unitary irreducible representations exist:
    \newline

    The ``principal continuous series" ${\mathcal{C}^\epsilon_h}$:
    Here $\mathbf{L}^2 < -{1\over 4}$ and we can write $h = \half + i\lambda$
    where $0 < \lambda \in \mathbf{R}$.  There is neither a lowest weight nor
    a highest weight state, and the weights are
    $m = \epsilon + n$ for $n \in \mathbf{Z}$.
    \newline

    The ``supplementary (or exceptional) continuous series" ${\mathcal{S}_h}$:
    Here $-{1\over 4} < \mathbf{L}^2 < 0$ and $0 < h < \half$ is real.  There is
    neither a lowest nor a highest weight state.  Only the case $\epsilon = 0$
    occurs and the weights
    are $m \in \mathbf{Z}$.
    \newline

    The ``discrete series" ${\mathcal{D}^\pm_h}$:
    Here
    $-{1\over 4} \le \mathbf{L}^2$ and $2h \in \mathbf{N}$.
    The highest weight representation $\mathcal{D}^+_h$ has weights $h + n$ for
    integer $n \ge 0$ and the lowest weight representation $\mathcal{D}^-_h$ has
    weights $- h - n$ for integer $n \ge 0$.
    \newline

    In addition to the unitary representations, there exists a
    series of finite-dimensional representations
    $\mathcal{D}_h$.  They are characterized by $h = 0, -\half, -1,
    \ldots$, and the weights are $h, h+1, \ldots, -h$.  These
    representations are not unitary with respect to any positive definite
    inner product, but, as we shall see below,
    they are (pseudo-) unitary with respect to an
    indefinite inner product on the state space.
    \newline

    We start by analyzing the
    coupling of the discrete series $\mathcal{D}^+_h$ with
    the finite-dimensional representations $\mathcal{D}_h$ (the
    analysis for $\mathcal{D}^+_h$ is entirely analogous and will not be
    presented here).

    \section{The discrete series $\mathcal{D}^+_h$}

    The highest weight representation $\mathcal{D}^+_h$ in the discrete
    series is generated from the highest weight state
    $\phi^{(h)}_h$
    satisfying $L_+ \phi^{(h)}_h = 0$, where
    $h \in {\half, 1, {3\over 2}, \ldots}$,
    by repeatedly applying $L_-$.
    The ladder of states
    \begin{equation}
        \phi_{h}^{(h)} \stackrel{L_-}{\longrightarrow} \phi_{h+1}^{(h)}
        \stackrel{L_-}{\longrightarrow}
        \phi_{h+2}^{(h)} \stackrel{L_-}{\longrightarrow} \cdots \label{LADDERSTATES}
    \end{equation}
    does not terminate.  Taking $e_h^{(h)} \equiv \phi_h^{(h)}$ to have unit
    normalization, the
    normalized basis
    \begin{equation}
        e_{h+k}^{(j)} \equiv {1\over \sqrt {k! \, (2h)\,(2h +
        1)\cdots(2h+k-1)}}
                \,\phi_{h + k}^{(h)}
    \end{equation}
    satisfies
    \begin{equation}
        \braket{e_{h+k}^{(h)}} {e_{h+m}^{(h)}} = \delta_{km},
        \label{INNERPROD}
    \end{equation}
    and
    \begin{eqnarray}
        L_-  \,e_m^{(h)}
                &= &\left\{ (-h + m + 1)\,(h +
                m)\right\}^{\half}\,e_{m+1}^{(h)},\qquad m = h, h+1, \dots
                \label{JPLUSACTION}
    \end{eqnarray}
    and
    \begin{eqnarray}
        L_+ \,e_{m+1}^{(h)} &= &\left\{ (-h + m + 1)\,(h +
                m)\right\}^{\half}\,e_{m}^{(h)},\qquad m = h, h+1, \dots
                \label{JMINUSACTION}
    \end{eqnarray}
    The representations $\mathcal{D}^-_h$ may be similarly
    constructed from a lowest weight state annihilated by $L_-$.

    \section{The finite-dimensional series $\mathcal{D}_h$}

    These representation $\mathcal{D}_h$, where
    $h \in {0, -\half, -1, -{3\over 2}, \ldots}$, may be generated from a highest weight
    state $\phi^{(h)}_h$, where $h\in \{0, -\half, -1, -{3\over 2},
    \ldots\}$.  The sequence of basis vectors
    \begin{equation}
        \phi_{h}^{(h)} \stackrel{L_-}{\longrightarrow} \phi_{h+1}^{(h)}
        \stackrel{L_-}{\longrightarrow}
         \cdots
         \stackrel{L_-}{\longrightarrow}
         \phi_{-h}^{(h)} \label{LADDERSTATES1}
    \end{equation}
    is finite.

    These representations can be made (pseudo-) unitary by
    choosing an indefinite inner product on the state space.
    To understand this, note that in a (pseudo-) unitary representation, the ladder
    operator $L_+$ is adjoint to $L_-$, so that we can
    write the inner product of the state $\phi_{h + k}^{(h)}$ with
    itself as follows
    \[
      \braket{\phi_{h + k}^{(h)}} {\phi_{h + k}^{(h)}} =
        \braket {(L_-)^k \phi_{h}^{(h)}}
        {(L_-)^k \phi_{h}^{(h)}}
        = \braket{\phi_{h}^{(h)}}
        {(L_+)^k(L_-)^k \phi_{h}^{(h)}}.
    \]
    Commuting the $L_+$ operators to the right, this becomes
    \begin{equation}
        \braket{\phi_{h + k}^{(h)}} {\phi_{h + k}^{(h)}}
        = k! \, (2h)\,(2h + 1)\cdots(2h+k-1)\,
        \braket{\phi_{h}^{(h)}}
        {\phi_{h}^{(h)}}.
    \end{equation}
    Since $h < 0$, we see that squared norms of
    the sequence of states in (\ref{LADDERSTATES1})
    have alternating signs, implying that
    the inner product is indefinite.  The normalized states
    \begin{equation}
        e_{h+k}^{(j)} \equiv {1\over i^k\sqrt {k! \, |2h|\,|2h +
        1|\cdots|2h+k-1|}}
                \,\phi_{h + k}^{(h)}
    \end{equation}
    have indefinite inner product
    \begin{equation}
        \braket{e_{h+k}^{(h)}} {e_{h+m}^{(h)}} = (-)^k\delta_{km},
        \label{INNERPRODINDEF}
    \end{equation}
    and the action of $L_\pm$ on this basis is given by
    \begin{eqnarray}
        L_-  \,e_m^{(h)} &= &i\,\left\{ \left(-h + m + 1\right)\,\left|h
                + m\right|\right\}^{\half}\,e_{m+1}^{(h)}
                \qquad m = h, h+1,\ldots, -h\nonumber\\
                &= &\phantom{i}\left\{ (-h + m + 1)\,(h +
                m)\right\}^{\half}\,e_{m+1}^{(h)},\qquad (-1)^{\half} \equiv
                +i
                \label{JPLUSACTION1}
    \end{eqnarray}
    and
    \begin{eqnarray}
        L_+ \,e_{m+1}^{(h)} &= &i\,\left\{ (-h + m + 1)\,|h +
                m|\right\}^{\half}\,e_{m}^{(h)}
                \qquad m+1 = h, h+1,\dots, -h\nonumber\\
                &= &\phantom{i}\left\{ (-h + m + 1)\,(h +
                m)\right\}^{\half}\,e_{m}^{(h)},\qquad (-1)^{\half} \equiv
                +i.
                \label{JMINUSACTION1}
    \end{eqnarray}
    These formulae are the continuation to negative $h$ of the
    corresponding formulae (\ref{JPLUSACTION}) and
    (\ref{JMINUSACTION}) for the discrete series.

    \section{Tensoring $\mathcal{D}^+_\half$ and $\mathcal{D}_{-\half}$}
    \label{HALFHALF}

    We now start our study of the coupling of the finite-dimensional
    representations of $SL(2,\mathbf{R})$ with the
    discrete series.

    As a warmup example, we study the product
    of the $h = \half$ discrete series representation with the $h = -\half$
    finite-dimensional representation.  The product representation
    will turn out to be reducible but not completely reducible
    (the representation matrices are not fully decomposable), and to contain
    null (zero norm) states.
    We will identify a
    natural cohomological reduction procedure that makes
    the product decomposable and eliminates the null states.
    In this particular example,
    after applying the reduction, the product of the two representations
    will give the trivial representation.  Schematically,
    indicating the cohomological reduction by an arrow, we have
    \[
        \mathcal{D}^+_\half\otimes\mathcal{D}_{-\half} \to
        \mathcal{D}_0 \equiv \mathbf{1}.
    \]
    To start our analysis, note that there is a highest weight state in the
    product state space given by
    \[
        \ket{0}\equiv
        e^{(\half)}_{\half}\otimes e^{(-\half)}_{-\half}.
    \]
    This state is annihilated by
    $L_+ \equiv L_+^{(\half)} + L_+^{({-\half})}$
    and satisfies
    \[
        \mathbf{L}^2 \, \ket{0} = 0,
    \]
    like the trivial representation.
    However, in contrast with the trivial representation, $L_- \ket{0}$ is not zero.
    Still, one can verify that $(L_-)^k \ket{0}$ has zero norm
    with respect to the indefinite inner product on the product
    space
    for $k = 1, 2, 3, \ldots$.  Concretely
    \[
        L_-^{n} \ket{0} = \sqrt {n}\,(n-1)! \, \ket{n},
    \]
    where the states
    \begin{eqnarray}
        \ket{n} &\equiv& \sqrt{n}\,\left(
                    e^{(\half)}_{n+\half}\otimes e^{(-\half)}_{-\half}
                    + i\,
                    e^{(\half)}_{n-\half}\otimes e^{(-\half)}_{\half}
                    \right)
    \end{eqnarray}
    are null.

    It would be nice if we could truncate the ladder of states
    generated by $L_-$ as soon as we reach a null state.  The
    proper way of doing this is by noticing that the Casimir
    operator $Q\equiv\mathbf{L}^2$ is nilpotent.  Therefore
    calculating its cohomology defines a natural reduction procedure on the state
    space.  The physicist will notice the analogy of this
    construction with BRST reduction, where $Q$
    is analogous to a BRST operator (see appendix \ref{BRST})

    To see that $\mathbf{L}^2$ is nilpotent,
    note that in terms of the basis consisting of $\ket{n}$ and the
    additional null states
    \begin{eqnarray}
        \ket{\tilde n} &\equiv& {1\over
        2\sqrt{n}}\left(-e^{(\half)}_{n + \half}\otimes
                            e^{(-\half)}_{-\half}
                    + i\,
                    e^{(\half)}_{n - \half}\otimes
                            e^{(-\half)}_{\half}\right),
                    \label{NTILDEN}
    \end{eqnarray}
    it is not hard to
    calculate
    \[
        \mathbf{L}^2 \ket{\tilde n} = \ket{n}, \qquad \mathbf{L}^2 \ket{n} = 0,
    \]
    leading to the matrix representation
    \begin{equation}
           \mathbf{L}^2 = \left(
                \begin{array}{cccc}
                        0 &  &  &  \\
                          & \fbox{\ensuremath{\begin{array}{rr}
                                    0 & 1 \\
                                    0 & 0
                                  \end{array}}
                            }
                             &  & \\
                          &  &
                            \fbox{\ensuremath{\begin{array}{rr}
                                    0 & 1  \\
                                    0 & 0
                                  \end{array}}
                            }
                                &  \\
                          &  &  &  \ddots
                \end{array}
            \right). \label{Q2}
    \end{equation}
    In this basis, the inner product is represented by the matrix
    \[
        G \equiv \left(
                \begin{array}{cccc}
                        1 &  &  &  \\
                          & \fbox{\ensuremath{\begin{array}{rr}
                                      & -1  \\
                                    -1 &
                                  \end{array}}
                            }
                             &  & \\
                          &  &
                            \fbox{\ensuremath{\begin{array}{rr}
                                      & -1  \\
                                    -1 &
                                  \end{array}}
                            }
                                &  \\
                          &  &  &  \ddots
                \end{array}
            \right)
    \]
    It is easily checked that $(\mathbf{L}^2)^2 = 0$ and that $\mathbf{L}^2$
    is hermitian
    with respect to the inner product $G$, as follows from
    \[
        \mathbf{L}^2 = (\mathbf{L}^2)^\dagger = G (\mathbf{L}^2)^+ G,
    \]
    where the dagger denotes the adjoint with respect to
    the indefinite inner product $\braket{\cdot}{\cdot}$ on the state space and
    the $+$ sign denotes the usual conjugate matrix.

    The set of states $\{\ket{n}\}$ defined above forms an $\mathbf{L}^2 = 0$
    subspace which is
    closed under $SL(2,\mathbf{R})$, but does not fall into any of the
    representations considered in section
    \ref{REPRESENTATIONS} unless we can get rid of
    the null states \[\{\ket{1}, \ket{2}, \ket{3}, \ldots\}.\]
    In addition, since the complementary set $\ket{\tilde n}$ is not
    closed under $SL(2,\mathbf{R})$, the full product representation is
    non-decomposable.

    The following procedure
    gets rid of these pathologies in the product.
    We construct the cohomology $\ker Q/\im Q$ with respect to the
    operator $Q = \mathbf{L}^2$.
    The cohomology consists of the single class $[\ket{0}]$,
    and is a one-dimensional, positive definite Hilbert space.
    Since the generators
    $L_i$ all commute with $Q$, they can consistently
    be reduced to the
    cohomology.
    The induced
    operators  $[L_+]$, $[L_-]$ and $[L_0]$ on the quotient space
    are defined as (\ref{AREDUCED})
    \[
      [L_i] \,[\ket{\phi}] = [L_i \ket{\phi}], \qquad \ket{\phi} \in \ker Q.
    \]
    and are indeed zero on the quotient space, corresponding to the trivial
    representation of $SL(2,\mathbf{R})$.

    \section{General product representations}

    We now generalize the method of section \ref{HALFHALF} to
    the analysis of a general product representation of a member of
    the discrete series and a member of the finite-dimensional series.

    In appendix \ref{J2ANALYSIS} we prove the result, important in
    what follows, that the Casimir operator $\mathbf{L}^2$ can be
    decomposed into Jordan blocks
    of dimension at most two (appendix \ref{INDEFINITESPACES}
    discusses properties of operators on indefinite inner product spaces).
    Since
    the one and two-dimensional Jordan blocks of $\mathbf{L}^2$ are of the
    forms $\left(\lambda_i\right)$ and
    \begin{equation}
        \left(
            \begin{array}{cc}
                \lambda_i & 1 \\
                0 & \lambda_i
            \end{array}
        \right), \label{J2BLOCK}
    \end{equation}
    we can build a nilpotent cohomology operator in the
    product space of two arbitrary representations in terms of
    $\mathbf{L}^2$ as the orthogonal direct sum
    \begin{equation}
        Q = Q_{\lambda_1} \oplus Q_{\lambda_2} \oplus \cdots,
        \label{Q}
    \end{equation}
    where $Q_{\lambda_i}$ is defined on the principal vector subspace
    $V_{\lambda_i}$ belonging to the eigenvalue $\lambda_i$ of $\mathbf{L}^2$ by
    \begin{equation}
            Q_{\lambda_i} = \mathbf{L}^2|_{V_{\lambda_i}} -
            \lambda_i.
        \label{QI}
    \end{equation}
    Since $Q_{\lambda_i}$ consists of one- and two-dimensional blocks of the
    forms
    $\left(0\right)$ and
    \begin{equation}
        \left(
            \begin{array}{cc}
                0 & 1 \\
                0 & 0
            \end{array}
        \right), \label{QBLOCK}
    \end{equation}
    $Q$ is nilpotent and is a valid cohomology operator.

    Taking the cohomology of $Q$ discards any Jordan
    blocks of dimension two in the decomposition of
    $\mathbf{L}^2$, so that all principal vectors of
    the reduced $[\mathbf{L}^2]$ will be eigenvectors.
    Equivalently, $[\mathbf{L}^2]$ can be decomposed into Jordan
    blocks of dimension $1$.

    Let us see how $Q$ affects the analysis of a
    general product representation.  As in the example of section
    \ref{HALFHALF}, negative $h$ representations may be
    present in the product in the form
    \begin{equation}
        \phi_{h}^{(h)}
        \Lmparrow \phi_{h+1}^{(h)}
        \Lmparrow \cdots
        \Lmparrow \phi_{-h}^{(h)}
        \Lmparrow
           \boxed{
            \begin{array}{c}
                \phi_{-h+1}^{(h)} \\
                \tilde\phi_{-h+1}^{(h)}
            \end{array}
           }
        \Lmparrow
            \boxed{
            \begin{array}{c}
                \phi_{-h+2}^{(h)} \\
                \tilde\phi_{-h+2}^{(h)}
            \end{array}
           }
        \Lmparrow
        \cdots
        \label{BADREP}
    \end{equation}
    The boxes represent two-dimensional subspaces spanned by null
    vectors $\phi_{-h+n}^{(h)}$ and $\tilde\phi_{-h+n}^{(h)}$ on
    which
    $\mathbf{L}^2$ has the Jordan normal form (\ref{J2BLOCK})
    and $Q$ has the form (\ref{QBLOCK}).  We see that just as in
    the example of section \ref{HALFHALF},
    the ladder of states generated from the highest weight $\phi_{h}^{(h)}$
    consists of null states for $m > -h$.

    Because of the additional states to the right of
    $\phi_{-h}^{(h)}$, this representation is not in the
    original class of irreducible
    positive or negative spin representations that we started
    out with.  In other words, the the original class of
    irreducible representations is not closed under
    multiplication, unless we can get rid of the extra states.

    Taking the cohomology with respect to $Q$ discards
    the boxed subspaces in the above diagram, and on the cohomology
    classes the representation has the familiar form
    \[
        [\phi_{h}^{(h)}]
        \Lmparrow [\phi_{h+1}^{(h)}]
        \Lmparrow \cdots
        \Lmparrow [\phi_{-h}^{(h)}].
    \]
    Henceforth, when we analyze product representations, it will
    always be assumed that we are working in the cohomology with respect to the
    associated operator $Q$.  This amounts to a redefinition of
    the product as the tensor product followed by the
    cohomological (BRST)
    reduction.

    \section{Characters}

    The analysis of the decomposition of product representations
    may be greatly simplified by reformulating it as an algebraic
    problem in terms of characters.

    We would like the expression for the character of a representation
    to be invariant under the above
    BRST reduction to the cohomology of $Q$.
    Since the discarded Jordan subspaces have zero metric
    signature, a definition for the character of a
    group element $U$ that will ignore these blocks is
    \begin{equation}
        \chi(U) = \sum_n \sig (V_{\lambda_n}) \lambda_n
        \label{CHARACTER}
    \end{equation}
    where $V_{\lambda_n}$ is the principal vector subspace
    corresponding to the eigenvalue $\lambda_n$ of $U$ and $\sig(\cdot)$
    denotes the signature.  Since the signature
    of a subspace is invariant under unitary transformations $V$ \cite{BOGNAR}, this
    gives a basis invariant expression
    invariant under conjugation $U \to VUV^{-1}$.

    At this point we just warn the reader that this
    definition needs modification in the
    infinite-dimensional case, where characters may only exist in the
    distributional sense.
    We will discuss this issue in
    more detail below.

    Using the following properties of the signature
    \begin{eqnarray}
        \sig (V \otimes W) &=& \sig V \cdot \sig W, \\
        \sig (V \oplus W) &=& \sig V + \sig W,
    \end{eqnarray}
    it follows that the characters satisfy the following important
    algebraic properties
    \begin{eqnarray}
        \chi (U_1 \otimes U_2) &=& \chi(U_1) \cdot\chi(U_2), \\
        \chi (U_1 \oplus U_2) &=& \chi(U_1) + \chi(U_2),
    \end{eqnarray}
    (where the right hand side of the first formula may be invalid in
    certain
    infinite-dimensional cases when the product of the
    distributions
    $\chi(U_1)$ and $\chi(U_2)$ may be undefined).
    These are exactly the properties that make the characters
    useful for analyzing the decomposition of a product
    representation into a direct sum of irreducible representations.
    The first property ensures that the character of a product
    representation is simply the product of the characters of the
    individual representations.  In other words,
    \[
        \chi^{R_1\otimes R_2} = \chi^{R_1} \chi^{R_2}.
    \]
    The second property then ensures that if $R_1\otimes R_2 = \sum_i
    n_i R_i$, then
    \[
        \chi^{R_1\otimes R_2} = \sum_i n_i \chi^{R_i},
    \]
    where the weight $n_i$ denotes the degeneracy of the representation $R_i$
    in the product.  In the indefinite metric case, the weight $n_i$ will
    be negative if the inner product on $R_i$ is of opposite sign to the
    usual conventions as in (\ref{INNERPROD}), or equivalently, if the
    lowest weight state has negative norm squared.
    This is due to the inclusion of the signature in the
    definition (\ref{CHARACTER}) above.

    We now discuss some issues specific to infinite-dimensional
    representations.
    Note that the definition (\ref{CHARACTER}) presupposes that the spectrum
    of the representation matrix $U$ is discrete and that the sum
    in (\ref{CHARACTER}) converges.  For our infinite-dimensional representations,
    this will not be true.  However, in this case, one can still define
    the characters of suitable smeared versions of the operators
    $U$ (see \cite{CHARACTERS}).  The characters become
    distributions defined on the group manifold.

    The group $SL(2,\mathbf{R})$ has three distinct families of conjugacy classes.
    These are the elliptic elements $\mathcal{E}$, indexed by the
    continuous parameter $\theta$, of which a typical element is
    given by $e^{i\theta L_0}$, and two families $\mathcal{H}_\pm$ of hyperbolic
    elements, each indexed by a continuous parameter $\sigma$ and
    with typical element given by $\pm e^{i\sigma L_1}$.

    The characters of the discrete series
    representation $\mathcal{D}^+_h$ were obtained in \cite{CHARACTERS}: The
    character of the elliptic element $e^{i\theta L_0}$ is
    \begin{eqnarray}
        \chi_{\mathcal{E}}^{(h)} (\theta)
         &= &{e^{ih\theta}\over 1 - e^{i\theta}} \nonumber\\
         &= &{-e^{i(h - \half)\theta}\over 2 i \sin \half
         \theta}. \label{CHARPOSITIVE}
    \end{eqnarray}
    and that of the hyperbolic elements is
    \begin{eqnarray}
        \chi_{\mathcal{H}_\pm}^{(h)} (\sigma) &=&
                \pm\chi_{\mathcal{E}}^{(h)} (\theta = i\sigma)
                \nonumber \\
                &=&
                 {\pm e^{-(h-\half)\sigma}\over 2 \sinh \half\sigma}.
    \end{eqnarray}
    We now compute the
    characters of the finite dimensional representations, which we denote
    by $\chi^{(h)}$ for $h \le 0$.
    Since these representations have the indefinite inner product
    (\ref{INNERPRODINDEF}), our definition (\ref{CHARACTER})
    does not reduce to the ordinary trace, but rather weighs the
    eigenvalues according to the metric signature of the corresponding
    eigenspace.
    We
    find, for $h < 0$
    \begin{eqnarray}
        \chi_{\mathcal{E}}^{(h)}(\theta) &=&  e^{ih\theta} \,
           (1 - e^{i\theta} + e^{2i\theta} - \cdots + (-)^{2h} e^{-2ih\theta})
              \nonumber\\
           &=& \left\{
                    \begin{array}{ll}
                         \displaystyle{\cos \left(h - \half\right)\theta \over \cos \half\theta}
                                & \qquad\mbox{if \ensuremath{2h} even,}          \\
                         \displaystyle{i \sin \left(h-\half\right)\theta\over\cos \half\theta}
                                & \qquad\mbox{if \ensuremath{2h} odd.}
                    \end{array}
                \right.
            \label{CHARNEGATIVE}
    \end{eqnarray}
    The characters of the hyperbolic elements are
    \begin{eqnarray}
        \chi_{\mathcal{H}_\pm}^{(h)}(\sigma) &=&
            \pm\chi_{\mathcal{E}}^{(h)}(\theta = i\sigma) \nonumber\\
            &=&
            \left\{
                    \begin{array}{ll}
                         \displaystyle{\pm\cosh \left(h - \half\right)\sigma \over \cosh \half\sigma}
                                & \qquad\mbox{if \ensuremath{2h} even,}          \\
                         \displaystyle{\mp\sinh \left(h-\half\right)\sigma\over\cosh \half\theta}
                                & \qquad\mbox{if \ensuremath{2h} odd.}
                    \end{array}
                \right.
    \end{eqnarray}

    \section{General products of discrete and finite series representations}

    The analysis of the product of discrete and finite series can
    now be carried out by calculating the algebra of the
    characters.
    This is most easily done by expanding the characters in
    powers of $e^{i\theta}$ or $e^{\sigma}$ as in the first line
    of
    (\ref{CHARNEGATIVE}), performing the multiplication and
    collecting terms.

    The following cases occur:
    \newline

    When $h_1>0$ (discrete series), $h_2 < 0$ (finite series)
    and $h_1 > |h_2|$, we have for the
    elliptic elements
    \begin{equation}
        \chi^{(h_1)}_{\mathcal{E}} \chi^{(h_2)}_{\mathcal{E}}
          = \chi^{(h_1 + h_2)}_{\mathcal{E}} - \chi^{(h_1+ h_2 + 1)}_{\mathcal{E}}
            + \cdots \pm \chi^{(h_1-h_2)}_{\mathcal{E}}.
    \end{equation}
    Since we saw that the characters of the hyperbolic elements are related to
    those of the elliptic elements by the analytic continuation
    $\chi^{(h)}_{\mathcal{H}_\pm}(\sigma) = \pm \chi^{(h_1)}_{\mathcal{E}}(\theta =
    i\sigma)$,
    it trivially follows that the same relation is satisfied by
    the hyperbolic characters.  Therefore
    \begin{equation}
        \mathcal{D}^+_{h_1} \otimes \mathcal{D}_{h_2}
         \to \sum_{h=h_1 + h_2}^{h_1 - h_2} (-)^{h_1 + h_2 - h}
         D^+_{h}, \label{DF1}
    \end{equation}
    where the arrow indicates cohomological reduction with respect to the operator
    $Q$.

    Note that this equation (\ref{FF})
    weighs representations with both positive and negative signs,
    which contain information about the signature of the inner product on the
    corresponding subspaces.  More precisely, one should read
    \[
        R_1 - R_2 \equiv R_1 \oplus (-R_2),
    \]
    where $-R_2$ denotes the representation $R_2$ but with
    opposite signature inner product.
    \newline

    When $h_1>0$ (discrete series), $h_2 < 0$ (finite series), $h_1 \le
    |h_2|$ and $h_1 + h_2$ is integral, then
    \begin{eqnarray}
        \chi^{(h_1)} \chi^{(h_2)}
          &=& \chi^{(h_1 + h_2)} + \chi^{(h_1+ h_2 + 1)}
            + \cdots + \chi^{(0)} \nonumber \\
            & & + \chi^{(-h_1 - h_2 + 2)} - \chi^{(-h_1 - h_2 + 3)}
            + \cdots \pm \chi^{(h_1 - h_2)},
    \end{eqnarray}
    where we have suppressed the subscript indicating
    elliptic/hyperbolic.  It follows that
    \begin{equation}
        \mathcal{D}^+_{h_1} \otimes \mathcal{D}_{h_2}
         \to \sum_{h=h_1 + h_2}^{0} D_{h}
            + \sum_{h = -h_1 - h_2 + 2}^{h_1 - h_2} (-)^{h + h_1 +
            h_2}D^+_h. \label{DF2}
    \end{equation}
    \newline

    When $h_1>0$ (discrete series), $h_2 < 0$ (finite series), $h_1 \le
    |h_2|$ and $h_1 + h_2$ is half-integral, then
    \begin{eqnarray}
        \chi^{(h_1)} \chi^{(h_2)}
          &=& \chi^{(h_1 + h_2)} + \chi^{(h_1+ h_2 + 1)}
            + \cdots + \chi^{(-\half)} + \chi^{(\half)}\nonumber \\
            & & - \chi^{(-h_1 - h_2 + 2)} + \chi^{(-h_1 - h_2 + 3)}
            - \cdots \pm \chi^{(h_1 - h_2)}.
    \end{eqnarray}
    It follows that
    \begin{equation}
        \mathcal{D}^+_{h_1} \otimes \mathcal{D}_{h_2}
         \to \sum_{h=h_1 + h_2}^{-\half} D_{h} + D^+_{\half}
            + \sum_{h = -h_1 - h_2 + 2}^{h_1 - h_2} (-)^{h + h_1 +
            h_2 + 1}D^+_h.  \label{DF3}
    \end{equation}
    \newline

    Finally, we calculate the coupling of finite series
    representations among themselves.  When $h_1 \le 0$ and $h_2 \le
    0$ (both finite series representations) and $|h_2| \le |h_1|$, then, again by
    expanding in powers of $e$ and collecting terms, we find
    \begin{eqnarray}
        \chi^{(h_1)} \chi^{(h_2)}
          &=& \chi^{(h_1 + h_2)} - \chi^{(h_1+ h_2 + 1)}
            + \cdots \pm \chi^{(h_1 - h_2)}.
    \end{eqnarray}
    It follows that
    \begin{equation}
        \mathcal{D}_{h_1} \otimes \mathcal{D}_{h_2}
         = \sum_{h=h_1 + h_2}^{-|h_1 - h_2|} (-)^{h_1 + h_2 - h}
         D_{h}.  \label{FF}
    \end{equation}
    In this case, no cohomological reduction is needed to obtain
    the right hand side.
    This decomposition is similar to what occurs in
    the representation theory of $SU(2)$.  This is as expected,
    since the finite dimensional representations of $SU(2)$ and
    $SL(2,\mathbf{R})$ are simply related by a rotation of two of the
    generators by $i$ (and renaming $-h = j$).
    However, unlike the $SU(2)$ case, the equation (\ref{FF})
    weighs representations with both positive and negative signs,
    which contain information about the signature of the inner product on the
    corresponding subspaces.
    \newline

    For completeness, we list the product decomposition of two
    discrete representations \cite{DECOMP}
    \begin{equation}
        \mathcal{D}^+_{h_1} \otimes \mathcal{D}^+_{h_2}
         = \sum_{h=h_1 + h_2}^\infty
         D^+_{h}. \label{DD}
    \end{equation}

    \section{Associativity}

    Let us denote by $R_1 \tilde\otimes R_2$ the above cohomological
    reduction of $R_1\otimes R_2$ with respect to the operator $Q$.
    The operation $\tilde \otimes$ is not associative, as can be
    seen from a simple counterexample
    \[
        \mathcal{D}^+_{\half}\tilde\otimes
          (\mathcal{D}^+_{\half}\tilde\otimes\mathcal{D}_{-\half})
          = \mathcal{D}^+_{\half}\tilde\otimes \mathbf{1} =
          \mathcal{D}^+_{\half},
    \]
    where we have used (\ref{DF2}),
    while
    \begin{eqnarray}
        (\mathcal{D}^+_{\half}\tilde\otimes
          \mathcal{D}^+_{\half})\tilde\otimes\mathcal{D}_{-\half}
           &=& (\mathcal{D}^+_{1} \oplus \mathcal{D}^+_{2} \oplus\mathcal{D}^+_{3}
                \oplus \cdots)
               \tilde\otimes
             \mathcal{D}_{-\half} \nonumber\\
             &=& \mathcal{D}^+_{\half} \oplus (-\mathcal{D}^+_{3\over 2})
             \oplus \mathcal{D}^+_{3\over 2} \oplus
             (-\mathcal{D}^+_{5\over 2}) \oplus \mathcal{D}^+_{{5\over 2}} \cdots
                \nonumber
    \end{eqnarray}
    where we have used (\ref{DD}) and (\ref{DF1}).

    From the example it is, however, obvious that the product can
    be made associative if we make the identification $R \oplus (-R) \sim 0$, where
    we remind the reader that $-R$ denotes the representation
    $R$ with inner product of opposite signature.
    In other words, we take the quotient with respect to
    sums of representations of the form $R \oplus (-R)$.  The definition
    (\ref{CHARACTER}) of the characters is obviously invariant
    with respect to this quotient, and the reduction of
    $\tilde\otimes$ with respect to the quotient is associative.

    \section{Coupling of finite-dimensional and continuous series representations}
    \label{couplcont}

    We will start by investigating the decomposition of
    products of the form ${\mathcal{C}^\epsilon_h} \otimes \mathcal{D}_{h'}$
    of a principal series and a finite-dimensional series representation.
    Here $h = \half + is$ for $s>0$ real, and $h' = -\half, -1, -{3\over 2},
    \dots$.

    The first case we consider is
    $$
        {\mathcal{C}^\epsilon_h} \otimes \mathcal{D}_{-\half}.
    $$
    With respect to the (non-normalized) bases $\ket{f_m^h}$ for $h = 1+is, -\half$,
    in which
    \begin{align*}
      L_- \ket{f_m^h} &= (m+1-h)\ket{f^h_{m+1}} \\
      L_+ \ket{f_{m+1}^h} &= (m+h) \ket{f_m^h},
    \end{align*}
    the Casimir decomposes into blocks of the form
    $$
      L^2 \longrightarrow \left(
           \begin{array}{cc}
             h(h-1) - m - {1\over 4} & (m+1-h) \\
             -(m+h)  & h(h-1) + m + {3\over 4}
           \end{array}
         \right)
    $$
    Diagonalizing, we find the eigenvalues of $L^2$ to be independent of
    $m$ and given by $\tilde h(\tilde h-1)$ for $\tilde h = h\pm
    \half$.  In other words, multiplication by $\mathcal{D}_{-\half}$
    takes us from a continuous series representation with $h = \half + is$
    to a direct sum of two representations with $\tilde h = (\half + is) \pm
    \half$.

    However, this is
    not a true decomposition since the inner product
    cannot be diagonalized simultaneously with $L^2$.  Indeed, it
    is a property of pseudo-hermitian operators such as $L^2$ that
    any complex eigenvalues come in conjugate pairs \cite{BOGNAR} and that
    the corresponding eigenspaces are null,
    and are not orthogonal but rather dual, meaning that with respect to the
    eigenbasis in each block the inner product takes the form
    \begin{equation}
      \left(
           \begin{array}{cc}
             0 & 1 \\
             1 & 0
           \end{array}
         \right)
         \label{sharpmetric}
    \end{equation}
    To denote this situation we write
    \begin{equation}
        {\mathcal{C}^\epsilon_h} \otimes \mathcal{D}_{-\half}
           = {\mathcal{C}^\epsilon_{h-\half}} \ \sharp\
             {\mathcal{C}^\epsilon_{h+\half}}  \label{sharp}
    \end{equation}
    where the representations
    ${\mathcal{C}^\epsilon_{h\pm\half}}$ with $h=\half + is \pm
    \half$ each has degenerate (null) inner product (which explains
    their absence from the traditional taxonomy of section 2).

    To calculate the coupling of ${\mathcal{C}^\epsilon_h}$ with
    an arbitrary finite representation $\mathcal{D}_{h'}$, we
    note that the latter can all be built up from products of $\mathcal{D}_{-\half}$.
    Indeed, from (\ref{FF}) we obtain the
    recursive relations ($h = -1, -{3\over 2} \dots)$
    \begin{equation}
        \mathcal{D}_{h}
         = \mathcal{D}_{h+\half} \otimes \mathcal{D}_{-\half}
           + \mathcal{D}_{h+1} \label{recurs}
    \end{equation}
    modulo the identification $R \oplus (-R) \sim 0$ discussed in
    the previous section.
    Together with (\ref{sharp}), this allows us to obtain the
    decomposition of an arbitrary product.  For example,
    using
    $$
     \mathcal{D}_{-1}
         = \mathcal{D}_{-\half} \otimes \mathcal{D}_{-\half}
           + \mathcal{D}_{0}
    $$
    we get, for $h = \half + is$,
    \begin{align*}
        {\mathcal{C}^\epsilon_h} \otimes \mathcal{D}_{-1}
        &= {\mathcal{C}^\epsilon_h} \otimes \mathcal{D}_{-\half}
          \otimes \mathcal{D}_{-\half} + {\mathcal{C}^\epsilon_h}
          \\
        &= \left( \mathcal{C}^\epsilon_{h+\half} \ \sharp\
                \mathcal{C}^\epsilon_{h-\half}
            \right) \otimes \mathcal{D}_{-\half}
             + {\mathcal{C}^\epsilon_h} \\
        &= \mathcal{C}^\epsilon_{h+1} \ \sharp\
                \mathcal{C}^\epsilon_{h-1} +
                \mathcal{C}^\epsilon_{h} \ \sharp\
                \mathcal{C}^\epsilon_{h}
                + {\mathcal{C}^\epsilon_h} \\
        &= \mathcal{C}^\epsilon_{h+1} \ \sharp\
                \mathcal{C}^\epsilon_{h-1}
                + {\mathcal{C}^\epsilon_h}
    \end{align*}
    where we have used $\mathcal{C}^\epsilon_{h} \ \sharp\
                \mathcal{C}^\epsilon_{h} = \mathcal{C}^\epsilon_{h}
                -
                \mathcal{C}^\epsilon_{h} \sim 0$, by
                diagonalization of the metric (\ref{sharpmetric}).

    Continuing in this vein, we obtain the product decompositions
    \begin{equation}
      {\mathcal{C}^\epsilon_{h_1}} \otimes \mathcal{D}_{h_2}
        = \sum_{k = 0}^{|h_2|-\half} c_{2|h_2|,k}
                \left( \mathcal{C}^\epsilon_{h_1+h_2+k} \ \sharp\
                \mathcal{C}^\epsilon_{h_1-h_2-k}\right)
    \end{equation}
    for $h_2 = -\half, -{3\over 2},
    \dots$, and
    \begin{equation}
      {\mathcal{C}^\epsilon_{h_1}} \otimes \mathcal{D}_{h_2}
        =  \sum_{k = 0}^{|h_2|} c_{2|h_2|,k}
                \left( \mathcal{C}^\epsilon_{h_1+h_2+k} \ \sharp\
                \mathcal{C}^\epsilon_{h_1-h_2-k}\right)
           + {\mathcal{C}^\epsilon_{h_1}}
    \end{equation}
    for $h_2 = 0, -1, -2, \dots$.  In these formulae,
    the nonzero entries in the
    table of coefficients $c_{ij}$
    are
    $$
      (c_{ij}) = \left(
       \begin{array}{ccccc}
         1 &   &   &   & \\
         1 &   &   &   & \\
         1 & 1 &   & &  \\
         1 & 3 &   & & \\
         1 & 5 & 1 & & \\
         1 & 7 & 9 & & \\
         1 & 9 & 21 & 1 & \\
         1 & 11 & 37 & 31 &\cdots \\
         \vdots & & & &
       \end{array}
      \right)
    $$
    Here each entry is generated from the sum of the three entries
    arranged in an L-shape above it.
    More formally, they are generated from the recursion
    \begin{align}
        c_{i+1,j+1} &= c_{i, j+1} + c_{i, j} + c_{i-1, j}  \\
        c_{0,j} &= 1 \\
        c_{2i, i} &= 1
    \end{align}
    which reminds one of the Fibonnacci sequence.  In fact, the actual
    Fibonnacci sequence makes an appearance in the next result:

    We now consider the decomposition of products of the supplementary continuous
    series ${\mathcal{S}_h}$, $0 < h < \half$ with the finite representations
    $\mathcal{D}_{h'}$.
    In fact, the allowed range of $h$ is a priori $0 < h < 1$, but
    representations indexed by $h$ and $1-h$ are isomorphic,
    allowing us to restrict attention to half the range.  As
    above, tensoring with $\mathcal{D}_{-\half}$ gives a direct
    sum of representations ${\mathcal{S}_{h+\half}}$ and
    ${\mathcal{S}_{h-\half}}$.  The first of these is isomorphic
    to ${\mathcal{S}_{\half-h}}$, while the second lies outside
    the allowed range for unitary ${\mathcal{S}_h}$ and once again denotes
    a representation with degenerate inner product.  Since the
    eigenvalues of $L^2$ are distinct and not conjugate, the
    corresponding eigenspaces are
    orthogonal, with the metric now taking the form
    \begin{equation}
      \left(
           \begin{array}{cc}
             0 & 0 \\
             0 & 1
           \end{array}
         \right).
    \end{equation}
    In contrast to the previous situation, the inner product factorizes
    and the representations are
    now independent.  If, in addition to
    the identification $R \oplus (-R) \sim 0$ of the
    previous section, we identify $R \oplus N \sim R$ whenever the
    inner product factorizes and $N$ is degenerate, we get the
    identity
    $$
      \mathcal{S}_h\otimes \mathcal{D}_{-\half} \to
        \mathcal{S}_{\half - h}.
    $$
    Generating $\mathcal{D}_{-n}$ by repeated tensoring with $\mathcal{D}_{-\half}$
    as in (\ref{recurs}), we can now calculate the general product
    decomposition
    \begin{equation}
      \mathcal{S}_h\otimes \mathcal{D}_{-k} \to
        \mathrm{fib}_k\,\mathcal{S}_{r^k(h)}.
    \end{equation}
    where $r^k \equiv r \circ r \circ \cdots \circ r$ denotes $k$
    applications of the reflection $h \mapsto \half - h$ and
    $\mathrm{fib_k}$ denotes the Fibonnacci sequence.

    \section{Products of two continuous series representations}
    \label{CONTPROBLEMS}

    The product decomposition of two continuous series
    representations is known and may be found in the references
    \cite{DECOMP}.  Here we point out an issue that
    should be addressed when combining those results with the formalism of the
    current paper.
    We have from reference \cite{DECOMP}
    \[
        \mathcal{C}^0_{h}\otimes\mathcal{D}^+_{\half}
         = \sum_{k = {3\over 2}}^\infty D^+_{k} + \int_{\half}^{\half
         +i\infty}
         dh'\, \mathcal{C}^{\half}_{h'}.
    \]
    We note the curious fact that the right hand side does not depend on
    $h$.  Consequently, neither could its product with the
    finite-dimensional series representation
    $\mathcal{D}_{-\half}$
    \[
      (\mathcal{C}^0_{h}\otimes\mathcal{D}^+_{\half})
      \tilde\otimes\mathcal{D}_{-\half}.
    \]
    On the other hand,
    \[
        \mathcal{C}^0_{h}\otimes(\mathcal{D}^+_{\half}
      \tilde\otimes\mathcal{D}_{-\half}) = \mathcal{C}^0_{h}
      \otimes \mathbf{1} = \mathcal{C}^0_{h},
    \]
    which does depend on $h$.  In other words, $\otimes$ does not
    associate over $\tilde\otimes$.  It would be interesting to
    investigate whether the definition of $\tilde \otimes$ can be
    extended to the case where both arguments are in the continuous
    series, in such a way that associativity is regained.  Since
    this question is likely to need technology beyond the scope
    of this paper \cite{DECOMP}, we defer it to future
    work.

    \section{Analytic continuation and  $SU(2)$}

    It is easily verified that that the finite-dimensional representations
    of the group $SL(2,\mathbf{R})$ may be
    related to the finite-dimensional representations of $SU(2)$ via
    the transformation $L_- = i J_+$, $L_+ = i J_-$, $L_0 = J_3$
    and $h = -j$.

    When we apply this transformation to the infinite-dimensional
    discrete series generators of $SL(2,\mathbf{R})$, we obtain an
    irreducible
    set of infinite-dimensional operators that formally
    satisfies the $su(2)$ Lie algebra.  These
    generators
    cannot be exponentiated to give a representation of the full $SU(2)$ group,
    since there exists a theorem
    stating that any continuous irreducible representation of a
    compact Lie group is finite-dimensional.  However, the generators can be
    exponentiated on a neighborhood of the identity, where they
    form so-called ``analytic" representations of $SU(2)$
    as defined by Segal \cite{SEGAL}.
    In \cite{MYSELF} and \cite{MYSELF1}, we studied the recoupling
    theory of these
    infinite-dimensional analytic representations of $SU(2)$.
    The results of the present paper are a
    straightforward adaptation of the methods used in
    \cite{MYSELF1}.

    In fact, the decomposition formulae (\ref{DF1}), (\ref{DF2}), (\ref{DF3}),
    (\ref{FF}) and (\ref{DD})
    correspond up to relative signs to the corresponding
    formulae involving the finite-dimensional (continuous) and
    infinite-dimensional negative spin (analytic) representations
    of $SU(2)$ studied in \cite{MYSELF1}.

    \section{Conclusion}

    In this paper we analyzed the decomposition of tensor products
    of the infinite dimensional (unitary)
    and the finite dimensional (non-unitary)
    representations of $SL(2,\mathbf{R}))$.
    Using classical results on indefinite inner product spaces,
    combined with cohomological methods, we were able to
    derive explicit decomposition formulae, true modulo a well-defined
    cohomological reduction, for the tensor products.

    As explained in section \ref{CONTPROBLEMS}, it would be
    interesting to revisit the existing results on tensor products
    between two continuous series representations to better understand
    whether an associative extension of $\tilde\otimes$ can be defined.

    It would also be interesting to determine whether some of
    these results can be generalized to other non-compact groups.

    \section*{Acknowledgments}

    I would like to thank the anonymous referee for contributing the
    idea developed in section \ref{couplcont}.
    I would also like to thank prof. Antal Jevicki and the Brown
    University Physics department for their support.

    \section*{Appendices}
    \appendix

    \section{BRST cohomology}
    \label{BRST}

    In the BRST formalism \cite{BRST},
    the analysis of physical states and
    operators is carried out in terms of an operator $Q$ that is hermitian
    and nilpotent.  In other words,
    \[
        Q^\dagger = Q, \qquad Q^2 = 0.
    \]
    States are called physical if they satisfy
    \[
        Q\ket{\phi} = 0,
    \]
    and are regarded as equivalent if they differ by a $Q$-exact
    state. In other words,
    \[
        \ket{\phi} \sim \ket{\phi} + Q \ket{\chi},
    \]
    where $\ket{\chi}$ is an arbitrary state.
    More formally, physical states are elements of the
    cohomology of $Q$, defined as the quotient vector space
    \[
        {\ker Q\, / \,\im Q}
    \]
    with elements
    \[
        [\ket{\phi}] \equiv \ket{\phi} + \im Q.
    \]
    The inner product on this quotient space may be defined in
    terms of the original inner product by
    noting that all elements of $\im Q$ are orthogonal to all
    elements of $\ker Q$, so that the induced inner product
    defined on equivalence classes in the cohomology by
    \[
        \braket {\phi + \im Q}{\phi' + \im Q} \equiv
        \braket{\phi}{\phi'}, \qquad \phi, \phi' \in \ker Q
    \]
    is well defined.

    A hermitian operator $A$ is regarded as physical if $[A, Q] =
    0$.  This ensures that $A$ leaves $\im Q$ invariant, so that
    the reduced operator $[A]$ defined on the cohomology classes by
    \begin{equation}
        [A]\,[\ket{\phi}] = [A\ket{\phi}]   \label{AREDUCED}
    \end{equation}
    is well-defined.

    \section{Operators on indefinite inner product spaces}
    \label{INDEFINITESPACES}

    We review a few
    general facts regarding hermitian operators on indefinite inner
    product spaces, also known as pseudo-hermitian operators \cite{BOGNAR}.

    It is important to be aware that not all results
    that are valid for positive definite spaces are valid when the
    inner product is not positive definite.  For example, not all
    pseudo-hermitian operators are diagonalizable.  A good
    counterexample is precisely the operator $Q = \mathbf{J}^2$ above.
    In addition, not all eigenvalues are necessarily real.  In
    particular, a pseudo-hermitian operator may have complex
    eigenvalues that come in conjugate pairs.

    For our purposes, we will restrict consideration to
    pseudo-hermitian operators with real eigenvalues, of which $\mathbf{L}^2$
    will be the relevant example.  The domain
    of such an operator $A$ can always be decomposed into a direct sum
    of pairwise orthogonal subspaces, in each of which
    we can choose a basis such that $A$ has the
    so-called Jordan normal block form
    \begin{equation}
        \left(
            \begin{array}{cccc}
                \lambda & 1       &        &        \\
                        & \lambda & 1      &        \\
                        &         & \ddots & \ddots \\
                        &         &        & \lambda
            \end{array}
        \right) \label{JORDAN}
    \end{equation}
    and the inner product has the form
    \begin{equation}
        \pm\left(
            \begin{array}{cccc}
                 &        &        & 1       \\
                &         & 1      &        \\
                & \cdots         &        &        \\
                1        &         &        &
            \end{array}
        \right). \label{GRAM}
    \end{equation}
    Notice that only the first vector in the subspace is an
    eigenvector of $A$ with eigenvalue $\lambda$.
    If the Jordan block has dimension
    larger than $1$, this eigenvector is null.
    The vectors $v$ in this subspace are
    called principal vectors belonging to the eigenvalue $\lambda$ and
    satisfy
    \[
        (A - \lambda)^m v = 0
    \]
    for some integer $m \ge 1$.  The sequence of vectors $v_i$ spanning
    this subspace satisfying
    \[
        A v_{i} = \lambda v_i + v_{i-1}
    \]
    is called a Jordan chain.

    \section{Jordan decomposition of $\mathbf{L}^2$}
    \label{J2ANALYSIS}

    In this appendix we prove that, in an arbitrary highest weight
    representation, $\mathbf{L}^2$ can be decomposed
    into Jordan blocks of dimension at most two.

    First, note that, since $\mathbf{L}^2$ commutes with $L_0$,
    we can decompose $\mathbf{L}^2$ into Jordan blocks in each
    eigenspace of $L_z$.  Consider such a Jordan block on a subspace
    $V_m$ consisting of
    principal vectors of $\mathbf{L}^2$ belonging to an eigenvalue
    $h(h-1)$ and with
    $L_0$ eigenvalue $m$.  Taking any vector $v$ in $V_i$,
    by applying $L_+$ to it repeatedly we will eventually obtain
    zero, since by assumption our representation is highest weight, so that
    the spectrum of weights of
    $L_0$ is bounded from below.
    This procedure gives a highest weight state $L_+^k v$, and we can
    use (\ref{CASIMIR}) to obtain $h(h-1)$ in terms of the $L_0$ eigenvalue
    $m - k$ of this
    highest weight state.  Since $m$ is integer or half-integer,
    the possible values of $h$ are also integer or
    half-integer, either positive or negative.  In the following,
    we shall take the positive solution $h>0$.

    Now consider the sequence of subspaces
    \[
        V_m \stackrel{L_+}{\longrightarrow} V_{m-1} \stackrel{L_+}{\longrightarrow}
        V_{m-2} \stackrel{L_+}{\longrightarrow} \cdots.
    \]
    Since $\mathbf{L}^2$ commutes with $L_+$, we see that
    $\mathbf{L}^2$ takes each of the subspaces $V_{i}$ to
    itself.  Furthermore, as long as $i \not\in \left\{-h + 1, h\right\}$, $L_+$
    cannot change the dimension of these subspaces since that
    would imply that $\ker L_-$ is not empty, so there would be
    highest weight states at values of $i$ inconsistent with $j$.
    In other words, the dimensions of the above sequence of spaces
    $V_i$ can at most jump at $i \in \left\{-h + 1, h\right\}$.
    As a corollary, taking into account the fact that the spectrum of
    $L_0$ is bounded from below, the above sequence terminates at either $i =
    -h + 1$ or $i = h$.

    Furthermore, $\mathbf{L}^2$ consists of a single
    Jordan block on each of the subspaces $V_i$.
    By assumption, this is true for the first element $V_m$
    of the sequence.  In general, assume that $\mathbf{L}^2$ consists of a
    single Jordan block on $V_i$ and consider $V_{i-1}$.  Since from (\ref{CASIMIR})
    we have that $L_- V_{i-1} = L_- L_+ V_i = (-\mathbf{L}^2 + L_0(L_0
    - 1) V_i$, we see that $L_- V_{i-1} \subseteq V_i$.  Now if
    $\mathbf{L}^2$
    were to consist of more than one Jordan block on $V_{i-1}$, each
    of these blocks would contain an eigenvector of $\mathbf{L}^2$.
    Since $L_-$ commutes with $\mathbf{L}^2$, all these
    eigenvectors will be taken by $L_-$ to eigenvectors in $V_i$,
    of which there is only one by assumption.
    Therefore $L_-$ is not one to one, its
    kernel on $V_{i-1}$ is nontrivial, and there is a highest weight
    state in $V_{i-1}$, which is inconsistent with $j$ unless
    $i \in \{-h + 1, h\}$.  This proves the assertion when
    $i \not\in \{-h + 1, h\}$.

    Now consider the case $i = h$.  The case $i = -h+1$ is similar.
    Suppose again that $\mathbf{L}^2$ consisted of more than one Jordan
    block, and therefore more than one eigenvector, on $V_{h-1}$.  This would imply,
    by the above argument,
    that the operator $L_-$ had
    nontrivial kernel on $V_{h-1}$.
    Choose $\tilde v_{h-1}\in V_{h}$ such
    that $L_- \tilde v_{h-1} = 0$.  Since $V_{h-1} = L_+ V_{h}$,
    there is a $\tilde v_{h}\in V_{h}$ such that $\tilde v_{h-1} = L_+ \tilde
    v_{h}$.  Then $0 = L_- L_+ \tilde v_{h} = (- \mathbf{L}^2 + L_0(L_0-1))
    \tilde v_{h}$, which implies that $\tilde v_{h}$ is an eigenvector of
    $\mathbf{L}^2$ and therefore is proportional to
    the unique eigenvector $v_{h}$ in $V_{h}$.
    Now note that $L_- v_{h-1}$ cannot be zero for more than one eigenvector in
    $V_{h-1}$, because if that were the case, then by the
    above argument there would be more than one linearly independent eigenvector of
    $\mathbf{L}^2$ in $V_{h}$.  Therefore we can find a $v_{h-1}$ such that
    $L_- v_{h-1} = v_{h}$.
    Then $L_+ L_- v_{h-1} = (-\mathbf{L}^2 + L_0 (L_0 + 1)) v_{h-1}
    = \left[-h(h-1) + (h-1)h\right] v_{h-1} = 0$,
    or $L_+ v_{h} = 0$.  But we had $0\ne\tilde v_{h-1} = L_+ \tilde
    v_{h}$, and $\tilde v_{h} \propto v_{h}$, which is a
    contradiction.  This proves the assertion when $i = h$.

    We have proved that $\mathbf{L}^2$ consists of a single Jordan
    block on each $V_i$.  This means that each $V_i$ contains one and only one
    eigenvector.  Consequently, since elements in the kernel of
    $L_+$ are automatically eigenvectors, the dimension of the
    $V_i$ can be reduced by at most one at each of the two transition
    points
    $V_{h} \stackrel{L_+}{\longrightarrow} V_{h-1}$ and
    $V_{-h + 1} \stackrel{L_+}{\longrightarrow} V_{-h}$.  Since the
    sequence terminates at either $V_{h}$ or $V_{-h+1}$, the
    initial space $V_m$ can be at most two-dimensional.  This
    completes the proof that the Jordan blocks of $\mathbf{L}^2$ are at most
    two-dimensional.


\begin{thebibliography}{99}
        \bibitem{SUONEONE}
            V. Bargmann, \textit {Ann. Math.} \textbf{48} (1947), 568; \newline
            A.O. Barut and C. Fronsdal, \textit{Proc. Roy. Soc. London}
            \textbf{A287} (1965), 532; \newline
            W.J. Holman and L.C.Biedenharn, Jr., \textit{Ann. Phys.}
            \textbf{39},
            (1966), 1; \newline
            I.M. Gel'fand, M.I. Graev, N.Ya. Vilenkin, \textit{Generalized
            functions, Volume 5. Integral Geometry and Representation
            Theory}, Academic Press, New York 1966; \newline
            L.C. Biedenharn, Jr., \textit{Ann. Phys.} \textbf{47}
            (1968), 205;\newline
            W. R\"uhl, \textit{The Lorentz Group and Harmonic
            Analysis}, W.A. Benjamin, Inc., New York 1970.
        \bibitem{BALASU}
            V. Balasubramanian, P. Kraus, A. Lawrence,
            \textit{Phys. Rev.} \textbf{D59} (1999) 046003.
        \bibitem{DECOMP}
            L. Puk\'anszky, \textit{Trans. Amer. Math. Soc.}
            \textbf{100} (1961), 116; \newline
            N. Mukunda, B. Radakrishnan, \textit{J. Math. Phys.}
            \textbf{15} (1974) 1320-1331; 1332-1342; 1643-1655;
            1656-1668; \newline
            J. Repka, \textit{Bull. Amer. Math. Soc.} \textbf{82}
            (1976) 930; \newline
            S. Molchanov, \textit{Math. USSR Izv.} \textbf{15}
            (1980), 113.
        \bibitem{FINTIMESINF}
            J. Lepansky, N.R. Wallach, \textit{Trans. Amer. Math. Soc.}
            \textbf{184} (1973) 223-246; \newline
            B. Kostant, \textit{J. Func. Anal.} \textbf{20} (1975)
            257-285; \newline
            C. Zuckerman, \textit {Ann. Math.} \textbf{106} (1977)
            295-308; \newline
            J.N. Bernstein, S.I. Gelfand, \textit{Compositio
            Math.}\textbf{41} (1980) 245-285.
        \bibitem{BOGNAR}
            J. Bognar, \textit{Indefinite Inner Product Spaces},
            Springer-Verlag, Berlin 1974;  \newline
            A.I. Mal'cev, \textit{Foundations of Linear Algebra},
            W.H. Freeman and Company 1963.
        \bibitem{CHARACTERS}
            I.M. Gel'fand and M.A. Naimark,
            \textit{I.M. Gel'fand - Collected Papers, Vol II}, p. 41,
            182,
            Springer-Verlag, Berlin-Heidelberg 1988; \newline
            S. Bal, K.V. Shajesh and D. Basu, \textit{J. Math. Phys.}
            \textbf{38} (1997) 3209; \newline
            D. Basu, S. Bal and K.V. Shajesh,
            \textit{J. Math. Phys.} \textbf{41} (2000) 461.
        \bibitem{SEGAL}
            I.E. Segal, \textit{Infinite-Dimensional Irreducible Representations
            of Compact Semi-Simple Groups}, Bull. Amer. Math. Soc.
            70 (1964), 155.
        \bibitem{MYSELF}
            A. van Tonder, \textit{Ghosts as Negative Spinors},
            Nucl. Phys. \textbf{B645} (2002) 371-386.
        \bibitem{MYSELF1}
            A. van Tonder, \textit{On the Representation Theory of Negative Spin},
            Nucl. Phys. \textbf{B645} (2002) 387-402.
        \bibitem{BRST}
            C. Becchi, A. Rouet and R. Stora, \textit{Phys. Lett.}
            \textbf{B52} (1974) 344;  \newline
            I.V. Tyutin, \textit{Lebedev preprint} FIAN \textbf{39} (1975),
            unpublished; \newline
            C. Becchi, A. Rouet and R. Stora, \textit{Ann. Phys.} \textbf{98} (1976),
            287;  \newline
            M. Henneaux, \textit {Phys. Rep.} \textbf{126} (1985), 1.
    \end{thebibliography}
\end{document}